\documentclass[a4paper,12pt]{article}

\usepackage{amsmath,amssymb,graphicx}
\usepackage{cite}

\numberwithin{equation}{section}

\newcommand{\ii}{\mathrm{i}}

\newcommand{\dd}{\mathrm{d}}

\newcommand{\e}{\mathrm{e}}

\newcommand{\tr}{\mathop{\mathrm{tr}}\nolimits}
\newcommand{\Tr}{\mathop{\mathrm{Tr}}\nolimits}

\newcommand{\ft}[2]{{\textstyle\frac{#1}{#2}}}

\begin{document}

\title{Matrix at slow roll: Nonrelativistic and Perturbative}%
\author{Corneliu Sochichiu\thanks{e-mail:
\texttt{sochichi@skku.edu}}\\
{\it  Dept. of Physics and University College}\\
{\it Sungkyunkwan University, Suwon 440-746, Korea}\\ and \\
{\it Institutul de
    Fizic\u a Aplicat\u a A\c S}\\
{\it  str. Academiei, nr. 5,
    Chi\c{s}in\u{a}u, MD2028 Moldova}
}%
%
%\subjclass{}%
%\keywords{}%

%\date{}%
%\dedicatory{}%
%\commby{}%
% ----------------------------------------------------------------
\maketitle
\begin{abstract}
 We analyze the slow roll limit of the massive version of time-dependent Yang--Mills type matrix model.  We find that this limit reproduces the one-loop  non-Hermitian matrix model, describing the dilatations of the local gauge invariant composite operators in the scalar sector of $\mathcal{N}=4$ super Yang--Mills theory. This coincidence is explained through the fact that the dialtation operator of the last can be identified with the radial time Hamiltonian. Further we explore the finite mass corrections to the slow roll action. The regular corrections in large mass expansions are coming from the modes with frequencies smaller than the mass parameter of the theory.
\end{abstract}
%\maketitle
%\tableofcontents
% ----------------------------------------------------------------
\section{Introduction}

Matrix models appear as a convenient tool for encoding complicated combinatorics in geometry related problems. IKKT and BFSS matrix models \cite{Ishibashi:1996xs,Banks:1996vh} were proposed as models describing IIB string with quantized worldsheet, and light-cone quantized membrane, respectively. These models are Yang--Mills type matrix models. They also proved to be relevant for background independent formulation of noncommutative Yang--Mills theory (see e.g. \cite{Sochichiu:2000ud-Sochichiu:2000rm}). 

Generally, one can obtain matrix mechanical models by the worldsheet quantization of a `sliced' description of extended objects \cite{Sochichiu:2000bg-Kiritsis:2002py-Sochichiu:2005ex,
Park:2008qe-Lee:2009ue-Lee:2010ey}. In the flat space-time such models a massless, but they may acquire a mass in a non-trivial background \cite{Berenstein:2002jq,Kim:2006wg}. Then the mass parameter corresponds to the strength of gravity background, and, according to `AdS/CFT dictionary', the large mass limit which means strongly coupled gravity and weakly coupled gauge theory. On the other hand, the large mass admits a slow roll expansion similar to nonrealtivistic limit. This expansion is the subject of this work

Massive Yang--Mills type matrix models also naturally arise in the radial time quantization \cite{Fubini:1973} of four-dimensional Yang--Mills \cite{Aharony:2005bq}.  As seen from the point of view of the real time formalism, the radial time Hamiltonian corresponds to the \emph{dilatation operator}. Therefore the large mass limit of these matrix models should correspond to the dilatation operator in the real time (see~\cite{Sochichiu:2007eu-Sochichiu:2008xd}) . Due to the scale transmutation, however, it is not clear whether the radial and real time quantizations are always equivalent. As we show below, even in the conformal theories, for which this is true, this mechanism imply nontrivial relations between the classical and one-loop level quantities.

The plan of the paper is as follows. In the next section we introduce the massive matrix theory and argue that the perturbation theory should be equivalent to the large mass expansion. Then, in the next section we take the slow roll limit of this theory and show that the resulting model can be associated to the dilatation operator in the scalar sector of $\mathcal{N}=4$ super Yang--Mills theory. Finally, we find the leading finite mass corrections to the slow roll action and conclude by the discussion of the results.

\section{Massive matrix theory}\label{sec:mass-mm}
Consider a purely bosonic massive matrix model described by the action,
\begin{equation}\label{mass-mm-action}
  S=\int\dd t\,\left(
  \frac{1}{2}\tr(\nabla_0 X_i)^2-\frac{1}{2}m^2 X_i^2+\frac{g^2}{4}\tr[X_i,X_j]^2
  \right).
\end{equation}
The indices $i,j$ run through $1,\dots,D$, and the covariant time derivative is defined by,
\begin{equation}\label{cov-d}
  \nabla_0 X_i=\dot{X}_i+[A,X_i],
\end{equation}
where $A$ is a non-dynamical U($N$) gauge field. The dot denotes the time derivative.
The system is invariant with respect to U($N$) time-dependent gauge transformations,
\begin{equation}
  X_i\to U^{-1}X_i U,\qquad A\to U^{-1}AU+U^{-1}\dot{U}.
\end{equation}

Let us consider the independent parameters controlling this model. A rescaling of the matrix variables $X_i\to g^{-\frac{1}{3}}X_i$ associated with the time rescaling $t\to g^{-\frac{2}{3}}t$ will set the coupling to $g=1$, changing at the same time the mass parameter $m\to m g^{-\frac{2}{3}}$. So the zero temperature theory effectively depends only on the combination $m^3/g^2$ of mass and coupling. Therefore, the large mass limit is equivalent to the weak coupling limit and \emph{vice versa}. In the case of a finite temperature, the above transformation modifies in addition the inverse temperature $\beta\to \beta g^{\frac{2}{3}}$.

The model described by the action \eqref{mass-mm-action} can be obtained, among others, as the bosonic part of the $\mathcal{N}=4$ supersymmetric Yang--Mills theory in the radial quantization \cite{Fubini:1973}.
In \cite{Berenstein:2005jq} this model was conjectured to provide (for $D=6$) a \emph{nonperturbative} description for dilatations in the BMN limit \cite{Berenstein:2002jq}. In contrast, in \cite{Bellucci:2004fh-Sochichiu:2006uz-Sochichiu:2006yv}, it was argued that it is a non-hermitian matrix model, which governs the dynamics associated to the dilatation operator at the one-loop level. In what follows we will consider the limit of slow dynamics for the action \eqref{mass-mm-action}, which is justified in the case of large mass, and show that the non-hermitian matrix model is reproduced in this limit. 

\section{The Slow roll limit}
Let us consider the slow roll expansion of the massive matrix model \eqref{mass-mm-action}.In order to do this, we express the matrix $X_i$ as a combination of `mostly forward' and `mostly backward' running matrices $\Phi_i$ and $\bar{\Phi}_i$, respectively,
\begin{equation}\label{pa-exp}
  X_i=\frac{1}{\sqrt{2m}}\left(\e^{-\ii m t}\Phi_i(t)+\e^{\ii m t}\bar{\Phi}_i(t)\right).
\end{equation}

The expansion \eqref{pa-exp} of the matrix $X_i$ is similar to the particle-antiparticle expansion in the non-relativistic limit in field theory.

To take the slow roll limit we substitute the decomposition \eqref{pa-exp} in the action \eqref{mass-mm-action} and take the limit of large mass, keeping the finite fixed value for,
\begin{equation}\label{gYM}
g^2_{\rm YM}=4\pi^2 (g/m)^2,
\end{equation}
 then dropping the terms vanishing in the $m\to\infty$ limit. As we aim the slow rolling limit we have also to discard the high frequency modes $\omega\gtrsim m$. This is equivalent to introducing UV cutoff $\Lambda\sim m$.
The resulting action takes the following form,
\begin{multline}\label{act-mm-SO6}
  S_{\rm slow}=
  \int \dd t\,
  \left(\tr
  \frac{\ii}{2}(\bar{\Phi}_i\nabla_0{\Phi}_a-\nabla_0
  {\bar{\Phi}}_i\Phi_i)\right.\\
  \left.+\frac{g_{\rm YM}^2}{16\pi^2}
  \left(\tr[\bar{\Phi}_i,\bar{\Phi}_j][\Phi_i,\Phi_j]+
  \frac{1}{2}\tr[\bar{\Phi}_i,\Phi_j][\bar{\Phi}_i,\Phi_j]
  \right)\right).
\end{multline}

This action coincides with the one corresponding to the dilatation operator for the scalar sector of $\mathcal{N}=4$ SYM constructed in \cite{Bellucci:2004fh-Sochichiu:2006uz-Sochichiu:2006yv}. The dilatation operator itself was first constructed in \cite{Beisert:2002bb-Beisert:2003jj}. (See also \cite{Sochichiu:2007am} for the construction of dilatation operators in the general case of renormalizable theories.)
This is not a simple accident and has an explanation in the conformal symmetry of $\mathcal{N}=4$ SYM. As a conformal theory, it can be equivalently quantized either in the real time framework or in the radial time formalism \cite{Fubini:1973}, in which the Hamiltonian of theory acquires an additional mass term corresponding to the engineering dimensions of the fields.

 The model described by the action \eqref{act-mm-SO6} is quite well understood: Its symmetries are well-known and the lowest energy eigenvalues and states can be found \emph{exactly} in the framework of \emph{spin bit model} \cite{Bellucci:2004ru-Bellucci:2004qx-Bellucci:2004dv}. Its thermodynamical analysis is similar to the one studied in \cite{Bellucci:2004fh-Sochichiu:2006uz-Sochichiu:2006yv}. 

Let us note, that since the effective coupling is $g^2/m^3\sim g_{\mathrm{YM}}^2/m$, the original matrix theory remains effectively perturbative for at least as far as $\lambda=g_{\mathrm{YM}}^2/4\pi^2 m$ is small, i.e. $g_{\mathrm{YM}}$ itself can be very large.

% --------------------------------------------------------------------
\section{`Relativistic' $1/m$ corrections}
The slow roll limit provided an easy way to retrieve otherwise more involved one-loop result of 
\cite{Beisert:2002bb-Beisert:2003jj,Sochichiu:2007am}. Inspired by this, it is interesting to investigate, whether we can gain more information on perturbative properties of the theory, following the same direction. Otherwise said, can we retrieve $1/m$ corrections to the slow roll action \eqref{act-mm-SO6} in the form of a systematic expansion? In the classical one-particle dynamics the answer to this question is easy: these corrections are casted in the energy expansion formula
\begin{equation*}
	E=\sqrt{m^2+p^2} \approx m+\ft{p^2}{2m}-\ft{p^4}{8m^3}+\dots  	
\end{equation*}
In the interacting quantum theory, as we will see, the situation is more tricky.

Indeed, to rich the slow roll action \eqref{act-mm-SO6} we discarded the following terms,
\begin{multline}\label{supressed}
	\Delta S_{\mathrm{non-slow}}=\int\dd t \biggl( \ft{1}{2m}\dot{\bar{\Phi}}_i \dot{\Phi}_i
	+\ft{1}{4m} \e^{-2\ii mt}\dot{\Phi}_i^2
	+\ft{1}{4m} \e^{2\ii mt}\dot{\bar{\Phi}}_i^2\\
	+ \ft{g_{\mathrm{YM}}^2}{16\pi^2}\e^{-2\ii mt}\tr [\Phi_i,\Phi_j][\Phi_i,\bar{\Phi}_j]
	+ \ft{g_{\mathrm{YM}}^2}{16\pi^2}\e^{2\ii mt}\tr [\bar{\Phi}_i,\bar{\Phi}_j][\bar{\Phi}_i,\Phi_j]\\
	+\ft{g_{\mathrm{YM}}^2}{64\pi^2}\e^{-4\ii mt}\tr [\Phi_i,\Phi_j]^2
	+\ft{g_{\mathrm{YM}}^2}{64\pi^2}\e^{-4\ii mt}\tr [\bar{\Phi}_i,\bar{\Phi}_j]^2\biggr).
\end{multline}
The terms in the first line of \eqref{supressed} are suppressed among others, because they are coming with a factor $\sim m^{-1}$. The remaining terms in the second and third lines come with the factors $\sim 1$. However, they have fast oscillating phase factors, therefore for slowly varying fields $\Phi$ and $\bar{\Phi}$ these terms also vanish, at least classically. The oscillating factors, however, realize a coupling between the slow and fast modes. Because of this, the situation becomes a little dangerous in the quantum theory: the coupling to higher frequencies can result in a finite or even large contribution to the slow roll action.

From the point of view of the original matrix theory \eqref{mass-mm-action}, the slow roll sector corresponds to the bunch of modes around the  free on-shell resonant modes $\omega=\pm m$. These modes, however, are coupled to modes with higher frequencies $|\omega|\gtrsim m$, as well as to the modes with lower frequencies: $|\omega|\lesssim m$. According to the general practice, in order to get the effective action for slow roll modes one should eliminate both higher and lower frequencies by integrating them, after introducing appropriate cutoffs.\footnote{Let us note, that since the expansion \eqref{pa-exp} is shifting the frequency of fields $\Phi$ with respect to the frequency of $X$, the notion of `high frequencies' and `low frequencies' are different for these two actions. Thus both `low frequency' and `high frequency' for the original action \eqref{mass-mm-action} will translate to `high frequency' of the reduced action \eqref{act-mm-SO6}. }

From a preliminary consideration it is clear, that integrating out the lower frequency modes will produce, as wanted, the expansion in positive powers of $1/m$. In contrast, the elimination of higher frequency modes would result in undesirable positive powers of mass. This singular contribution appears because the mass parameter in the slow roll limit itself plays the role of a UV cut off and the introduction of a new UV cut off  produces a redefinition of the scale. Therefore, to keep $m$ as the UV cut off scale, we have to discard any frequency above this scale.

So let us evaluate the contribution coming from the frequencies  inside the interval $\pm m$ . To do this, let us split the matrix field $X_i$ in slow rolling mode given by \eqref{pa-exp} and low frequency mode correction $u_i$ as follows,
\begin{equation}
	 X_i=\frac{1}{\sqrt{2m}}\left(\e^{-\ii m t}\Phi_i(t)+\e^{\ii m t}\bar{\Phi}_i(t)\right)
	+ u_i(t),
\end{equation}
where $u_i(t)$ is containing only low frequency modes with $|\omega|\lesssim\Lambda\ll m$. Plugging this expansion into the classical action \eqref{mass-mm-action}, we get in the leading order in $u_i$,
\begin{multline}
	S=S_{\mathrm{slow}}+\int\dd t\, \biggl(\ft{1}{2m} \nabla_0\bar{\Phi}_i \nabla_0\Phi_i\\
	+
	\ft{1}{2}( \nabla_0 u_i)^2-\ft{1}{2}m^2u_i^2\\
	+ \ft{g^2}{2m} \bigl(
	-u_j[\bar{\Phi}_i,[\Phi_i,u_j]]
	+2 u_i[\bar{\Phi}_j,[\Phi_i,u_j]]-u_i[\bar{\Phi}_i,[\Phi_j,u_j]]\\
	-u_j[{\Phi}_i,[\bar{\Phi}_i,u_j]]+2 u_i[{\Phi}_j,[\bar{\Phi}_i,u_j]]-u_i[{\Phi}_i,[\bar{\Phi}_j,u_j]]\bigr)\biggr)+\dots,
\end{multline}
where the dots denote higher order terms. The first term in the integral is the regular $1/m$ correction which we discussed above. Another corrections come from the integration of $u_i$, so we can write,
\begin{equation}\label{DScorr}
	\Delta S_{\mathrm{corr}}\approx\int\dd t\, \biggl(\ft{1}{2m} \nabla_0\bar{\Phi}_i \nabla_0\Phi_i\biggr)
	-\ft{1}{2} \log\det\bigl(m^2+\nabla_0^2+\ft{g^2}{m} \mathcal{M}\bigr),
\end{equation}
where $\mathcal{M}$ is a local operator defined defined by the action,
\begin{multline}
	(\mathcal{M}\cdot u)_i=
	-[\bar{\Phi}_k,[\Phi_k,u_i]]
	+2[\bar{\Phi}_j,[\Phi_i,u_j]]
	-[\bar{\Phi}_i,[\Phi_j,u_j]]\\
	-[{\Phi}_k,[\bar{\Phi}_k,u_i]]
	+2 [{\Phi}_j,[\bar{\Phi}_i,u_j]]
	-[{\Phi}_i,[\bar{\Phi}_j,u_j]],
\end{multline}
in components,
\begin{multline}
	\mathcal{M}_{jab'}^{ia'b}=\ft{1}{2}\bigl( 
	\bar{\Phi}_{ja}^{~c}\Phi_{ic}^{~a'}\delta_{b'}^{b}
	-\bar{\Phi}_{ja}^{~a'}\Phi_{ib'}^{~b}
	-\bar{\Phi}_{jb'}^{~b}\Phi_{ia}^{~a'}
	+\bar{\Phi}_{jc}^{~b}\Phi_{ib'}^{~c}\delta_{a}^{a'}\\
	\bar{\Phi}_{ic}^{~a'}\Phi_{ja}^{~c}\delta_{b'}^{b}
	-\bar{\Phi}_{ib'}^{~b}\Phi_{ja}^{~a'}
	-\bar{\Phi}_{ia}^{~a'}\Phi_{jb'}^{~b}
	+\bar{\Phi}_{ib'}^{~c}\Phi_{jc}^{~b}\delta_{a}^{a'}\bigr).
\end{multline}

Hence, the leading term in the expansion of the log of the determinant is,
\begin{equation}\label{logdet}
	-\ft{1}{2} \log\det\bigl(m^2+\nabla_0^2+\ft{g^2}{m} \mathcal{M}\bigr)\approx
	-\ft{g^2}{2m^3}\int\dd t\Tr \mathcal{M} -\ft{1}{2m^2}\int\dd t\Tr \mathcal{A}+\dots, 
\end{equation}
where we dropped the constant terms the dots stand for the higher orders. The operator $\mathcal{A}$ is a remnant of the covariant derivative,
\begin{equation}
	\mathcal{A}\cdot u=[A,[A,u]],\quad 
	\mathcal{A}_{ab'}^{a'b}=
	A_{a}^{~c}A_{c}^{~a'}\delta_{b'}^{b}
	-2 A_{a}^{~a'}A_{b'}^{~b}
	+A_{c}^{~b}A_{b'}^{~c}\delta_{a}^{a'}.
\end{equation}
\emph{A priori} we should keep this term as well, because there is no kinetic term for the gauge field $A$, and as a result no frequency suppression occurs for it.

Evaluating the traces, we can rewrite the leading corrections to the slow roll action \eqref{DScorr} in the form as follows,
\begin{multline}\label{corr-final}
	\Delta S_{\mathrm{corr}}\approx 
	\int\dd t\, \biggl(\ft{1}{2m} \nabla_0\bar{\Phi}_i \nabla_0\Phi_i
	+\ft{g^2(D-1)}{2m^3}\bigl(N\tr \bar{\Phi}_k \Phi_k-\tr \bar{\Phi}_k \tr \Phi_k \bigr)\\
	+\ft{1}{m^2}\bigl(N\tr A^2-(\tr A)^2\bigr)\biggr).
\end{multline}

Let us have a closer look on the corrections \eqref{corr-final}. The effect of the third term in \eqref{corr-final} is to modify the measure of the non-dynamical gauge field. This term only modifies the gauge fixing prescription (if any) and has no dynamical implication on the theory. The second term has the form of a mass deformation for the non-abelian part of the matrix field. From the point of view of the theory of anomalous dimensions, this term corresponds to a modification of the letter's bare dimensions, a finite wave function renormalization. As this term gives the engineering dimension, this contribution will not change the theory, at least for closed sectors of the dilatation operators.

Let us turn to the first term. The first term is the `classical' correction to the slow roll action. It modifies the sypmplectic structure of the model. This modification is serious since it seems to double the number of degrees of freedom. It happens because the terms of the type $\dot{\Phi}^2$ and $\dot{\bar{\Phi}}^2$ are cut off in the slow roll limit. As a result two perturbative vacua  $\omega=+m$ and $\omega=-m$ became disjoint and slow fluctuations around these vacua will appear as independent degrees of freedom. Of course, inclusion of higher order corrections will make these vacua unstable and the redundancy in the degrees of freedom will be eliminated.

\section{Discussion}
In this paper we considered the slow roll limit of a massive matrix theory. In particular, it was shown that this limit reproduces the model for anomalous dimensions of local gauge invariant composite operators in the scalar sector of $\mathcal{N}=4$ supersymmetric Yang--Mills theory. This is explained by the coincidence of the perturbative dilatation operator with the radial time Hamiltonian in the slow roll limit. Even so, this is remarkable since it implies that the result of one-loop calculations can be easily obtained by a limiting procedure. Inspired by this we performed the calculation of  leading $1/m$ corrections to the slow roll action. The most nontrivial result of these corrections is the doubling of the number of degrees of freedom resulted from the separation of particles and antiparticles by an impenetrable barrier in the slow roll limit. This phenomenon is common for low energy effective theories.
 
Another point worth of being mentioned is the relation to the integrability. As it was shown in  \cite{Bellucci:2004fh-Sochichiu:2006uz-Sochichiu:2006yv} the planar limit of the perturbative non-Hermitian matrix model \eqref{act-mm-SO6} is equivalent to an \emph{integrable} model consisting of a collection independent Heisenberg spin chains. The correspondence just shown in this paper implies that massive matrix model is \emph{planar integrable} too in the large mass limit. It is interesting to check if this planar integrability extends to finite mass as well. So far, the correction  \eqref{corr-final} spoils the symplectic structure of the slow roll action, so it is unclear whether the planar integrability is preserved or not. 

\subsection*{Acknowledgments}
I benefited from useful discussions with O-Kab Kwon and Jeong-Hyuck Park. I am also grateful to Shinsuke Kawai for his critical remarks, as well as for useful discussions. 

This work was supported by Korean NRF grant no. 2010-0007637.

% ----------------------------------------------------------------
\bibliographystyle{JHEP}
\bibliography{nrmm}
\providecommand{\href}[2]{#2}\begingroup\raggedright\endgroup

\end{document}